\documentstyle[12pt,epsf,epsfig]{article}     

\def\lsim{\lower0.6ex\vbox{\hbox{$ \buildrel{\textstyle 
<}\over{\sim}\ $}}}
\def\rsim{\lower0.6ex\vbox{\hbox{$ \buildrel{\textstyle 
>}\over{\sim}\ $}}}
\def\beq{\begin{equation}}
\def\eeq{\end{equation}}
\def\beginapjbib{\begingroup \section*{\large \bf References}
         \parskip=.5ex plus 1.0pt
         \def\bibitem{\par \noindent \hangindent\parindent
                \hangafter=1}}
\def\endapjbib{\par \endgroup}
\def\alwaysmath#1{{\ifmmode{#1}\else{$#1$}\fi}}
\def\he#1{\hbox{\alwaysmath{{}^{#1}}{\rm He}}}

\def\li#1{\hbox{\alwaysmath{{}^{#1}}{\rm Li}}}
\def\sun{${\,_\odot}$}

\def\hii{H\thinspace{$\scriptstyle{\rm II}$}~}
\def\etal{{\it et al.}~}
\begin{document}
\begin{flushright}
UMN-TH-1514/96\\
OSU-TA-24/96  \\
astro-ph/9611166 \\
November 1996
\end{flushright}
\vskip 0.75in
 
\begin{center} 

{\Large{\bf   The Primordial Abundance of \he4: An Update}}
 
\vskip 0.5in
{  Keith~A.~Olive$^1$, Evan~Skillman$^1$  \\ and \\ Gary~Steigman$^{2,3} $}
 
\vskip 0.25in
{\it
$^1${School of Physics and Astronomy,
University of Minnesota, Minneapolis, MN 55455}\\
 
$^2${Department of Physics,
The Ohio State University, Columbus, OH 43210}\\
 
$^3${Department of Astronomy, The Ohio State University, Columbus, 
OH 43210}\\

}


\vskip .5in
{\bf Abstract}
\end{center}
\baselineskip=18pt 
 
~We include new data in an updated analysis of helium in low 
metallicity extragalactic \hii regions with the goal of deriving 
the primordial abundance of \he4 ($Y_{\rm P}$).   We show that the 
new observations of Izotov \etal (ITL) are consistent with previous 
data.  However they should not be taken in isolation to determine 
$Y_{\rm P}$ due to the lack of sufficiently low metallicity points. 
We use the extant data in a semi-empirical approach to 
bounding the size of possible systematic uncertainties in the 
determination of $Y_{\rm P}$.  Our best estimate for the primordial 
abundance of \he4 assuming a linear relation between \he4 and O/H is 
$Y_{\rm P} = 0.230 \pm 0.003$(stat) based on the subset of \hii regions 
with the lowest metallicity; for our full data set we find $Y_{\rm P} = 
0.234 \pm 0.002$(stat).  Both values are entirely consistent with our 
previous results.  We discuss the implications of these values for 
standard big bang nucleosynthesis (SBBN), particularly in the context 
of recent measurements of deuterium in high redshift, low metallicity 
QSO absorption-line systems.

\newpage
\baselineskip=18pt 
\noindent
\section{Introduction}

Determining the primordial abundances of the light elements D, 
\he3, \he4, and \li7 is crucial for testing the standard model 
of big bang nucleosynthesis (SBBN) and in using SBBN to set 
constraints on cosmology (e.g., the baryon density) and on 
particle properties (e.g., the number of light degrees of 
freedom contributing to the energy density of the early universe 
(cf. Walker \etal 1991)).  Each of the light nuclides poses 
different problems in the quest for its primordial abundance.  
For example, deuterium is destroyed during the pre-main sequence 
evolution  of stars.   As such, the relation between its observed 
and primordial abundances depends on the amount of material which 
has been cycled through stars (i.e., on Galactic chemical 
evolution).  Uncertainties in Galactic evolution currently 
dominate any attempt to infer the primordial abundance of 
deuterium from local (``here and now") observations although 
such data do provide a lower bound to its primordial value.
  
The situation for \he3 is much worse (i.e., even more subject 
to uncertainties in stellar and Galactic evolution) since stars 
of all masses destroy some -- but not all -- of their prestellar 
\he3 and low mass stars are expected to be net producers of \he3.  
Thus, to infer the primordial abundance of \he3 from observations 
``here and now" requires that the balance between destruction, 
survival and new production be known to an accuracy that eludes 
us at present.  Since \li7 is observed in extremely metal-poor 
stars, Galactic chemical evolution plays virtually no role in 
using the data to infer its primordial abundance.  However, since 
the very metal-poor stars are also very old, they have had a long 
time to modify their surface abundance of \li7 from its nearly 
primordial initial value.  Uncertain corrections for stellar 
evolution dominate the uncertainty in the derived primordial 
abundance of lithium.  

Since \he4 is the most abundant nuclide in 
the universe (after hydrogen) it may be observed throughout the 
universe (not only ``here and now'') more accurately than any of 
the other nuclides. For individual, careful observations, 
the abundance of \he4 may be determined to 5\% or better 
(0.012 in Y, the \he4 mass fraction).  Since its primordial 
abundance is expected to be quite large ($\sim 25\%$ by mass) 
and \he4 has been observed in the less processed, metal-poor, 
extragalactic \hii regions, the extrapolation to primordial 
is minimal (differences in Y $\lsim$ 0.002 -- 0.004). However, 
in the context of SBBN, the predicted primordial abundance 
of \he4 is rather insensitive to (only logarithmically dependent 
on) the one free SBBN parameter -- the nucleon-to-photon ratio 
$\eta~(\eta \equiv n_N/n_{\gamma}; \eta_{10} \equiv 10^{10} \eta$).  
For this reason it is necessary to establish the primordial 
abundance of \he4 very accurately.  Particular attention must be 
paid to possible systematic uncertainties as well as to the usual 
statistical errors.

Although, compared to the other light isotopes, the 
primordial \he4 abundance is rather insensitive to 
chemical and stellar evolution, corrections for such 
effects are not entirely absent.  In the course of 
their evolution stars burn hydrogen to helium and 
when they die they return this processed material 
(containing new \he4 along with heavy elements (``metals'') 
such as C, N, O) to the interstellar medium (ISM) 
polluting the primordial $^4$He.  To minimize the 
contribution from stellar-produced $^4$He, the best 
observational targets are those regions whose heavy 
element abundances are low, suggesting the least 
contamination from stellar and galactic chemical evolution. 
 This has led virtually all investigators to concentrate on 
the low metallicity, extragalactic \hii regions (Searle \& 
Sargent 1971; Peimbert \& Torres-Peimbert 1974; Lequeux et al. 
1979; Kunth \& Sargent 1983; Torres-Peimbert et al. 1989; 
Pagel et al. 1992, hereafter PSTE; Skillman \& Kennicutt 1993; 
Skillman et al. 1994, 1996, hereafter S; Izotov, Thuan, \& 
Lipovetsky 1994, 1996 hereafter ITL).  However, since in even 
the lowest metallicity regions observed (with metallicity as low 
as 1/50 of solar) some \he4 was produced along with the heavy 
elements, Peimbert \& Torres-Peimbert (1974) proposed to correlate 
$Y_{\rm OBS}$ with metallicity and to extrapolate to zero metallicity 
in order to infer $Y_{\rm P}$.  Since the heavy element mass fraction, 
$Z$, is not observed, the abundances of oxygen and/or nitrogen have 
usually served as surrogates (e.g., $Z \approx$ 20(O/H)).

A previous analysis (Olive \& Steigman 1995, hereafter OS), 
considered the \he4, O/H and N/H data of PTSE and of S, for 
49 separate, low metallicity, extragalactic \hii regions. In that 
analysis the correlation between O/H and N/H was explored, and it 
was concluded that the nitrogen observed in these objects is 
dominated by a primary contribution (i.e., N/H increasing linearly 
with O/H) with a small but not entirely insignificant secondary 
component.  The fits of Y versus O/H or Y versus N/H were not 
sensitive to the inclusion or exclusion of galaxies which show 
Wolf-Rayet features or of galaxies which deviated from the mean 
N/H versus O/H relation.  There was, however, a small difference 
between the values of $Y_{\rm P}$ derived from the fits of $Y$ 
versus O/H and $Y$ versus N/H.  Given that there are both primary 
and secondary contributions to nitrogen, this is not unexpected 
(Fields 1996); we will return to this issue below.  Overall, it 
was found that the data were well described by a linear fit of 
$Y$ to O/H with an intercept at zero metallicity of $Y_{\rm P} = 
0.232 \pm 0.003$.  In addition to the above statistical error, 
various contributions were described which might lead to an overall 
systematic uncertainty of order $\pm$ 0.005 (see also PSTE).

In the past year or two the \he4 abundance, key to the consistency 
of SBBN, has come under great scrutiny.   Most recently, ITL 
presented new data which they claimed provides evidence for 
$Y_{\rm P}$ in excess of 0.24.  In this paper, we consider the 
ITL data and ask if they are consistent with those of PSTE and S.  
We conclude that they are and we propose an explanation 
for the apparent contradiction.  We then use all extant data 
(PSTE, S and ITL) to derive the current best estimates 
for the primordial abundance of \he4.  We also attempt 
to use the data in a semi-empirical approach to estimating 
the size of the possible systematic uncertainty in $Y_{\rm P}$.

\section{The Old Data used in OS}

OS used the data of PSTE and S for 49 separate extragalactic 
\hii regions.  In minimizing the extrapolation to zero metallicity, 
the lowest metallicity \hii regions play a crucial role.  As a result 
a ``first cut'' was made in OS eliminating those \hii regions (albeit 
only 8 out of 49) with N/H $\geq 1.0 \times 10^{-5}$ and O/H $\geq 1.5 
\times 10^{-4}$.  Note that all of the \hii regions retained are 
metal-poor compared to the Sun where (O/H)\sun\ = $8.5 \times 10^{-4}$ 
and (N/H)\sun\ = $1.1 \times 10^{-4}$.  Nevertheless, the OS ``first cut" 
data set spans one order of magnitude in oxygen abundance ($15~ \lsim 
10^6 (O/H)~ \lsim 150$) and a factor of $\sim 25$ in nitrogen abundance 
($4 ~\lsim 10^7 (N/H)~ \lsim 100$).  OS also considered an even more 
metal-poor subset (``second cut"), retaining the 21 (out of 41) \hii 
regions with O/H $\leq 8  \times 10^{-5}$ ([O/H] $\leq -1$).  This 
more restricted 2nd cut set still has a modest dynamical range in its 
oxygen and nitrogen abundances:  $ 15 ~\lsim 10^6$(O/H) $\lsim 80$ and 
$ 4 ~\lsim 10^7$(N/H)$ ~\lsim 40$.

As mentioned above, OS investigated the correlation 
between N/H and O/H for these \hii regions.  Although the 
variation of nitrogen with oxygen is of interest for the 
study of chemical evolution, it must be emphasized that 
the evolution of the very low mass host galaxies of these extragalactic 
\hii regions is likely dominated by local -- in space 
and in time -- processes.  Different \hii regions may be 
``caught" at different evolutionary epochs (e.g., just
before or just after a starburst, shortly before or 
immediately after a supernova explosion, etc.).  
Overall, OS found a strong correlation between N and O, 
and that at low metallicity nitrogen is predominantly 
primary (varying linearly with oxygen) with a small, 
but not insignificant, secondary component (proportional 
to the square of the oxygen abundance). This can be seen 
from a power law fit to the data, N/H $\propto$ (O/H)$^\alpha$, 
where OS found $\alpha = 1.31 \pm 0.07$.  This behavior is 
confirmed with our enlarged data set, now containing 62 
distinct extragalactic \hii regions (labelled set B below), 
for which we find $\alpha = 1.21 \pm 0.06$.  Alternatively, 
the predominantly primary nature of nitrogen can be seen by 
fitting the data with a linear N/O versus O/H relation for 
which we find: 
\beq 
N/O = (0.023 \pm 0.002) + (76 \pm 19) (O/H)
\eeq
Note that the ``primary" component dominates for O/H 
$\lsim$ 3.1 $\times$ $10^{-4}$, and that for our entire ``first 
cut" range the ``secondary" to ``primary" ratio varies from 
5 -- 50 \%. This is in agreement with Pagel \& Kazlauskas (1992) 
who concluded that ``primary" nitrogen dominates at low metallicity.

Pagel, Terlevich \& Melnick (1986) noted that \hii regions which 
showed Wolf-Rayet spectral features often had larger abundances of 
both helium and nitrogen compared to \hii regions with the same 
oxygen abundance but lacking such features.  OS searched for such 
an effect but found no statistically significant correlation, so OS 
did not exclude any \hii regions with Wolf-Rayet features and neither
will we in our analysis here.  This conclusion is supported by ITL
and by Kobulnicky \& Skillman (1996). 

In the previous Y versus O/H analysis, OS found for the 1st cut 
(2nd cut) set $Y_{\rm P} = 0.232 \pm 0.003$ (0.229 $\pm$ 0.005).  
Since then the data on extragalactic \hii regions has increased 
significantly.  Izotov et al. (1994) presented observations of 10 
\hii regions, four of which overlap those in the S set used in OS.  
These new data were incorporated in the analysis of Olive \& Scully 
(1996) who found that Y$_{\rm P}$ derived from Y versus O/H for the 
expanded 1st cut set (now containing 47 \hii regions) is slightly 
higher (although within 1$\sigma$ of OS): $Y_{\rm P} = 0.234 \pm 0.003$.

\section{The New Data of ITL}

In their most recent work Izotov et al. (1996) have data from 
observations of 28 new \hii regions.  Of these one region is 
included in their 1994 set and four others are contained in the 
PTSE set, including a reobservation of IZw18.  Thus we now have data 
for 78 distinct extragalactic \hii regions (several of which have 
been observed by two or more independent groups).  In the following 
analyses we impose the same low metallicity 1st cut restriction as 
in OS, eliminating the same 8 regions from the PTSE set.  Thus the 
largest, low metallicity set available for analysis consists of 70 
regions (set A).  However, according to several criteria ITL exclude 
10 regions (including IZw18) from their analysis; here we accept their 
judgment and with one notable exception (see below) eliminate the 
same regions from consideration.  Since eight of the ITL excluded 
regions are not contained in the PSTE or S sets, our data set is 
now reduced to 62 distinct \hii regions (set B).  For our second cut 
(in this case at O/H $< 8.5 \times 10^{-4}$; [O/H] $< -1$) we have 
32 distinct \hii regions (set C).  However, before proceeding it is 
necessary to consider the relation between the ITL data and those of 
PTSE and S.  Are they consistent? 

   One issue concerns the calculation of the statistical errors reported 
with the observed line ratios (corrected for reddening).  PSTE report 
errors derived from the total counts in the line and the continuum, and 
terms accounting for the sky subtraction and the read noise of the 
detector. Errors were not calculated for the reddening correction, the 
flat field correction, or the wavelength-dependence of the sensitivity 
(the ``flux" calibration), but care was taken such that these errors were 
of order, or smaller, than the calculated errors (see Simonson 1990).
Skillman et al.\ (1994) include all of these terms in their uncertainties 
(see e.g., equation 2 in Skillman et al.\ 1994).

  ITL do not provide sufficient details to permit us to determine 
how the errors in their emission line ratios are calculated.  But 
they do state that their spectra are ``in excellent agreement" with 
those previously published in the literature.  Frequently, their 
brighter lines are quoted with errors between 0.1 and 0.2\% .  This 
is to be contrasted with the analyses of PSTE and S where the minimum
uncertainties usually lie in the range of 1 to 2\% .  It is interesting
that ITL do comment that the residuals in their flux calibration curve 
are `` $\le$ 5\% ."  From this statement it appears possible that ITL
might have underestimated the uncertainties in some of their reported 
line ratios by at least a factor of 25 (this may be a lower limit which 
could increase if other errors, e.g., the reddening correction, are 
accounted for).  The possibility that ITL may have significantly
underestimated their observational errors is supported by, for example, 
the  5$\sigma$ change in the reported oxygen abundance for 0940$+$544N 
from 7.37 $\pm$ 0.02 in ITL94 to 7.48 $\pm$ 0.01 in ITL96.

  There are further reasons to suspect that ITL may have
underestimated their errors.  For example, it is not possible 
to know relative line ratios with an accuracy better than the
calibration of the telescope/spectrograph/detector
combination.  In assembling a set of spectrophotometric
standard stars for use with the HST, Oke (1990) used CCDs
that were measured to be linear to within 0.2\% and found
that standard star measurements were repeatable to ``about
1\% over most of the spectral range and a little larger in
the ultraviolet and near infrared."  Thus, it would be prudent
to adopt 1\% as a reasonable lower limit to the error on any 
measured emission line ratio.

   It is beyond the scope of this paper to devise our own corrections
to the error estimates of ITL.  Indeed, in some cases the line ratio 
errors may not be very far off (e.g., for the weakest lines, the quoted 
errors become comparable to and even surpass the uncertainty in the 
flux calibration).  Thus, for our analysis we simply adopt the ITL
reported uncertainties.  

ITL have adopted an approach of using the data for four HeI lines 
($\lambda$4471, $\lambda$5876, $\lambda$6678, $\lambda$7065) 
in a self-consistent analysis whose goal it is to determine 
simultaneously the recombination and the collisional excitation 
contributions to the observed emissivities.  By insisting that the 
line ratios have their recombination values after correction for 
collisions, ITL determine the electron densities self-consistently.  
The virtue of this approach is that it avoids the use of uncertain 
electron densities determined indirectly from [SII].  The problem 
with this approach is its reliance on the $\lambda$7065 line which, 
although sensitive to collisional excitation (albeit with an uncertain 
collision strength), is well-known (Robbins 1968; PSTE; G. Ferland, 
Private Communication) to be subject to fluorescence.  The observed 
$\lambda$7065 line strengths may well be providing a measure of the 
optical depth through the \hii regions rather than of the effect of 
collisional excitation (although ITL argue to the contrary).  
Unquantified radiative transfer effects, complicated by unknown \hii 
region geometry, dust/gas, etc., may introduce large uncertainties in 
the ITL approach which call into question the efficacy of their reliance 
on this line.  Further, their approach requires that ITL have good data 
for all four lines and this forces them to reject otherwise good 
observations of \hii regions when they have sufficiently accurate 
data for only two or three of the four lines.  In contrast to the ITL 
approach, neither PSTE nor S use the $\lambda$7065 line in their analyses 
and Peimbert (1996) notes that for the relatively low electron densities 
common to \hii regions the collisonal correction is usually quite small. 

  ITL use their method to analyze their data in a number of different
ways.  They have compared the He abundances derived from the observed
emission line strengths using the independent sets of recombination
line emissivities by Brocklehurst (1972; hereafter B72) and Smits
(1996; hereafter S96).  For $\lambda\lambda$ 5876, 4471, and 6678,
the new S96 emissivities are in good agreement with those of B72
(see Table 3 of S96 for a comparison).  Over the relevant range of
electron temperature and density, the B72 and S96 emissivities for
$\lambda$5876 and $\lambda$6678 agree to within one percent and
for $\lambda$4471 to within two percent.  There is roughly a 40\%
difference for $\lambda$7065 where B72 was in error (see discussion
in Smits 1991a,b).  Given the error in the B72 $\lambda$7065 emissivities,
it does not make sense for ITL to use the B72 emissivities in concert
with their method.

  ITL also use two different sets of collisional excitation rates
to correct for the contribution of emission from collisional excitation
from the metastable 2$^3$S level of He I.  Clegg (1987; hereafter C)
calculated these rates from the 19-state (up to n $=$ 4) R-matrix
computation by Berrington \& Kingston (1987).  Kingdon \& Ferland
(1995; hereafter KF) have calculated new rates, based on the 29-state
(up to n$=$ 5) computation of Sawey \& Berrington (1993).  Figure 1
compares the results of the C and KF calculations for the relative
rates of collisionally excited emission to recombination emission
(C/R) for the four He lines used by ITL.  Note the excellent
agreement for $\lambda$5876 and $\lambda$6678.  The small change in
the rate for $\lambda$4471 is due to both a change in the rate to
the n $=$ 4 level, and to the addition by KF of rates to
two n $=$ 5 levels.  The large change in the $\lambda$7065 C/R
value is mainly due to the difference between the recombination
rates.  C used B72 for these rates and KF used S96 for these rates.
Since the emissivities are implicit in the C/R calculations, it
makes no sense to combine C with S96 nor KF with B72 as ITL
have done.  While the differences for $\lambda$5876 and $\lambda$6678
will be negligible, the differences for $\lambda$4471 will be
significant, and for $\lambda$7065 very large.

 Therefore, when using the ITL analysis, the recombination emissivities 
must be restricted to those from S96 while any analysis which avoids 
the $\lambda$7065 line may use either B72 or S96.  Since previous 
analyses (PSTE; S) have avoided this line (and have employed B72/C), 
it is interesting to compare the Y values derived from the ITL 
observations using S96/KF (their ``best" combination) to those 
derived using B72 for the recombination emissivities, C for the 
collision strengths, along with the electron densities determined from 
[SII].  The ITL data reveal these differences to be quite small, in 
fact a weighted average of difference of the \he4 mass fractions, 
$\langle$ Y(S96/KF) - Y(B72/C/SII) $\rangle = -0.003$ which is much 
less than the typical errors in the individual Y determinations.  Indeed, 
most of this difference is traceable to the $\approx$ 1\% differences 
between the KF and C collision strengths which, however, lead to no 
significant difference (i.e., $\lsim$ 0.2\%) when ITL evaluate Y$_{\rm P}$ 
using S96/KF or S96/C (see Table 7 of ITL 1996).  Within the uncertainties, 
Y(S96/KF) = Y(B72/SII).  This agreement suggests that the analyses of 
PSTE and of S have not been biased by their reliance on B72, and that 
their results using B72/C/SII should be directly comparable to those of ITL.

In contrast, ITL found from linear regressions of Y on O/H using their own 
data Y$_{\rm P}$(S96/KF) = 0.243 $\pm$ 0.004 whereas OS, using PSTE and S 
data, found Y$_{\rm P}$(B72/SII) = 0.232 $\pm$ 0.003.  Given the relatively 
small statistical uncertainties such a large difference suggests an 
inconsistency between PSTE and S on the one hand and ITL on the other.  
Some have interpreted this apparent discrepancy as an indicator of the 
true size of the systematic errors in Y$_{\rm P}$ determinations and have 
embraced the larger ITL values as a better probe of Y$_{\rm P}$.  ITL 
apparently believe the inconsistency is real and they claim it is 
traceable to the use by PSTE and S of the older B72 emissivities.  But we 
have just demonstrated, using the ITL data, that this cannot be the case; 
within the errors for all ITL \hii regions: Y(S96/KF) = Y(S96/C) = Y(B72/SII).  
Therefore, we must be able to compare the value of Y$_{\rm P}$ derived from 
the ITL observations using the B72/C/SII combination with that found by OS 
using the PSTE and S data.  For the data from the ITL preferred set of 27 
\hii regions, a linear regression of Y on O/H yields Y$_{\rm P}$ = 0.241 
$\pm$ 0.004, still quite high compared to the OS result.

Why, then, do ITL find such a large value for $Y_{\rm P}$ (0.241) 
compared to that found by OS from the data of PSTE and of S (0.232)?  
We believe it is due to the absence of the very lowest metallicity 
\hii regions from the ITL set after they have excluded selected regions 
from their analysis.  In Figure 2 we show (with the error bars suppressed) 
the Y versus O/H data used in OS (open circles) along with the new 
ITL data (filled circles).  The crossed circles are the 10 \hii regions 
that ITL excluded from their analyses.  Note that where there is overlap 
in O/H, the ITL Y values are intermingled with those from PSTE and S.  
Indeed, there are six \hii regions in common between ITL and OS (not 
counting those ITL regions, shown in Figure 2, which ITL excluded from 
their fits) with four having higher and two with lower Y values than 
their PSTE or S counterparts.   It is indeed surprising that ITL and OS 
find such significantly different values for Y$_{\rm P}$.  However note 
that the ITL data set does not extend to as low an oxygen abundance as 
the set employed by OS.  This more limited range in metallicity for the 
ITL data set gives them less leverage in determining the slope of the
Y versus O/H relation.  Indeed, in the Y versus O/H fit of the ITL data 
set, the slope is found to be 64 $\pm$ 48.  This is reminiscent of the 
earlier work of Kunth \& Sargent (1983) whose limited metallicity range 
also led to an indeterminate slope.  Kunth \& Sargent (1983) made the 
appropriate choice based on their data and took a weighted mean of their 
data to determine Y$_{\rm P}$, finding a high value of 0.245 $\pm$ 0.003 
which is very similar to those found by ITL.  Indeed, when ITL used S96/KF 
or S96/C they found slopes consistent with zero which suggests that their 
Y$_{\rm P}$ estimates are effectively weighted means.  

To test this hypothesis, we have refit the Y versus O/H relation for 
the OS set excluding the four lowest O/H points.  For this modified 
``1st cut" OS set of 37 \hii regions we find $Y_{\rm P} = 0.237 \pm 
0.004$, significantly higher than the previous result for the full 
``1st cut" set (0.232 $\pm$ 0.003).  This reflects the high weights 
in the fits carried by the lowest metallicity \hii regions which tend 
to have low Y values with small uncertainties.  It is this modified OS 
value of 0.237 which should be compared to the ITL result of 0.241.  
Within the statistical uncertainty they are entirely consistent.  To 
explore this further, we have employed the modified ``1st cut" fit to 
describe the ITL data.  The reduced $\chi^{2}$ for this fit is 0.50 
(to be compared with 0.44 for ITL's own fit).  Based on the F-test 
(Bevington \& Robinson 1992) there is a 39\% chance that their data is 
drawn from a distribution described by our fit.  In addition, we have 
used the modified set of 37 \hii regions in a ``statistical bootstrap" 
of 40,000 runs (see Olive \& Scully 1995) and we found that $Y_{\rm P}$ 
exceeded 0.241 13\% of the time. This is shown in Figure 3.  These tests 
lead us to conclude that the new ITL data do not differ statistically 
from the older PSTE and S data used by OS. 

With this as justification, we proceed to analyze the combined data of 
PSTE, S and ITL.  In this analysis we adopt the ITL data derived 
using the electron densities determined from SII so that we may 
have an internally consistent data set\footnote{We have used the 
values of Y and $\sigma_Y$ given by ITL.  We note that the error 
in Y was not statistically propagated from that in the 
abundance by number, $y$.  Thus the quoted errors are somewhat 
larger (by $\sim 30$\%) than they should be.  We 
did not correct for this.}.  Although we avoid using the ten \hii 
regions discarded by ITL, eight of them are rather insignificant in 
the sense that since their abundances have large uncertainties (and 
another is at intermediate metallicity), they would have low weight in 
our fits and not much influence on our derived value of $Y_{\rm P}$.  
The one exception is IZw18 which provides the lowest metallicity 
point and for which ITL have good data.  ITL exclude IZw18 from 
their analysis because the $\lambda$5578 line is subject to 
absorption by interstellar sodium.  Not being able to use all four 
HeI lines in their self-consistent approach, ITL eliminate IZw18 
from their analysis.  Izotov (Private Communication) has kindly 
provided us with the $^4$He abundance derived from their data using 
the $\lambda$6678 line and the [SII] electron density and we use this 
information to include this data point in our subsequent analysis.
Note that when IZw18 is included with the other 27 ITL data points, 
their intercept drops to 0.239 $\pm 0.004$ (the effect is minor since 
the error is relatively large -- in OS the error in IZw18 is diminished 
due to multiple as well as high quality observations).  
If we make our second cut to the ITL data 
including IZw18, the intercept drops to 0.231 $\pm$ 0.006.  This result 
is now completely consistent with the OS value of 0.229 $\pm$0.005.

\section{Results}

\subsection{$Y_{\rm P}$ From The Helium-Metallicity Correlation}

We adopt several approaches to using the \hii region data (set B with 
62 separate \hii regions and the lower metallicity set C with 32 
regions) to infer the primordial abundance of $^4$He, $Y_{\rm P}$.  
Since primordial helium has been contaminated with the debris of 
stellar ejecta, the most common approach has been to use the 
metallicity information to probe the correlation of Y with
Z (either O/H or N/H) and to extrapolate this empirical relation to 
zero metallicity to find $Y_{\rm P}$ (Peimbert \& Torres-Peimbert 
1974).  With increasing numbers of very low metallicity \hii regions, 
this extrapolation is quantitatively quite small ($\Delta{Y} \approx$ 
0.002 -- 0.004).  In OS we showed that the extant data do, indeed, 
justify a positive correlation between Y and O/H (N/H).  To explore 
this from a somewhat different perspective, consider the following:  
For set B we have computed the weighted mean of the helium 
abundances, $\langle Y \rangle$ and in Figure 4 we plot the 
residuals, Y $- \langle Y \rangle$ as a function of the oxygen 
abundance.  At low metallicity almost all the residuals are 
negative, while the positive residuals appear only at higher 
metallicity.  Thus a one parameter fit to the Y versus O/H data 
fails to account for the clear helium -- oxygen correlation and is a 
poor fit to the data.  Therefore we next try to fit the data with linear 
Y versus O/H (N/H) regressions.  These two-parameter fits describe the 
data very well (see Table 1).  Note that if instead of our new ``first cut" 
set B we had used the data for all 70 independent \hii regions (set A), 
there would be no difference in our derived value of $Y_{\rm P}$.  
Similarly, there is no difference exceeding 0.001 in $Y_{\rm P}$ if we 
exclude the ITL value (Izotov, Private Communication) for IZw18 from our 
fits.

{}From Table 1 we notice that the values of $Y_{\rm P}$ derived from 
the Y versus N/H relations are systematically higher (but only by 
$\lsim 1\sigma$) than those inferred from Y versus O/H.  This effect 
is also present in OS and is entirely to be expected on the basis of 
the primary/secondary origin of nitrogen (Fields 1996). As Fields 
(1996) shows, the primary/secondary origin for nitrogen, compared to 
the primary origin for oxygen implies that when $Y_{\rm P}$ is derived 
from a {\em linear} correlation with N/H the result will exceed the 
``true" value derived from the {\em linear} Y versus O/H relation.  
The quantitative difference between the two regressions will depend 
on the details of chemical evolution models as well as on the observed 
correlation of N with O and will be explored in future work (Fields, 
Olive \& Steigman 1996).  For this reason we adopt for our estimates 
of 
$Y_{\rm P}$ those values derived from the Y versus O/H regressions 
for 
the B(C) sets (Table 1),
\beq
Y_{\rm P} = 0.234 \pm 0.002 (0.230 \pm 0.003)
\eeq
for which $Y_{\rm P}^{2\sigma} \leq 0.239 (0.237)$.  If we assume 
(PTSE) that the metallicity Z and O/H may be related by Z $\approx 
20(O/H)$, the set B(C) Y versus O/H fits imply: $\Delta{Y}/\Delta{Z} 
\approx 6 (12)$, consistent with the steep slopes found by PSTE,
Olive, Steigman, \& Walker (1991) and OS.

\vskip .5in
\begin{table}[h]
\centerline {\sc{\underline{Table 1:} Linear Fits for $Y$ versus. 
$O/H$}}

\vspace {0.1in}
\begin{center}
\begin{tabular}{|cccccc|}                     \hline \hline
Set &  \# Regions  & ${\chi}^2/dof$ &  $Y_{\rm P}$  &  $10^{-2}
 \times$ slope &$Y_{\rm P}^{2\sigma}$ \\ \hline

O/H: B & 62 & 0.58 & $.234 \pm .002$ & $ 1.25 \pm 0.27$& 
0.239 \\ 
O/H: C & 32  & 0.59 & $.230 \pm .003$ & $2.31 \pm 0.65$ & 
0.237\\
N/H: B & 62 & 0.61 &$.237 \pm .002$ & $25.7 \pm 5.9$& 
0.241 \\
N/H: C & 32  & 0.63 & $.232 \pm .003$ & $61.5 \pm 17.5$ & 
0.238 \\
\hline
\end{tabular}
\end{center}
\end{table}

\subsection{$Y_{\rm P}$ From A Few Good \hii Regions}

For the fits described above, the extrapolation from the lowest 
metallicity \hii regions to zero metallicity is minimal ($\Delta{Y} 
\approx 0.002 - 0.004$).  Nonetheless, it is true that any 
extrapolation to zero metallicity could be avoided since the helium 
abundance inferred from the observations of any one \hii region (with 
non-zero metallicity) should provide an upper limit to $Y_{\rm P}$.  
For a very low metallicity \hii region such an upper limit may even 
provide a reasonable estimate of $Y_{\rm P}$.  In this context, IZw18, 
the most metal-poor \hii region, is an ideal candidate (Kunth \etal 1994) 
since it has been the subject of careful study by several independent 
groups (PSTE, Skillman \& Kennicutt (1993) and ITL).  A weighted mean of 
the five observations of the two separate knots in IZw18 yields,

\beq
Y(IZw18) = 0.230 \pm 0.004
\eeq
with a $2\sigma$ upper bound of 0.237.  In terms of statistical 
accuracy this result is fully competitive with the value of 
$Y_{\rm P}$ derived in the previous section from 62(32) \hii 
regions.  

Of course, it should be kept in mind that the abundance inferred for 
any one \hii region might be anomalous.  Therefore the value of Y 
derived from the average of several \hii regions is also of interest.  
In such an analysis, as more regions are included, the mean value 
(weighted) of Y will increase, but if the errors are statistical, the 
error in the mean will decrease.  As a result, for $N$ \hii regions the 
one-(or two-)$\sigma$ upper bound to Y will first decrease with $N$, 
then level off and, as $N$ is further increased, it will eventually 
increase monotonically.  This behavior is seen in Figure 5 where we 
show the weighted means, and the 2$\sigma$ bounds to the 
weighted 
means of Y derived from the $N$ lowest helium abundance \hii 
regions.  
Note that for $2 \leq N \leq 13$, the mean varies from 0.229 to 
0.231 
while for $2 \leq N \leq 14$, $\langle Y \rangle \leq 
0.236$(2$\sigma$).  
It is not unreasonable to infer from these results that,
\beq
Y_{\rm P} \leq 0.230\pm 0.003
\eeq
with, $Y_{P}^{2\sigma} \leq 0.236$.  If, instead, we take the weighted
means of the regions with the lowest values of O/H, we 
obtain a similar result.  This illustrates the potentially great 
value of very careful analyses of a handful of the lowest metallicity 
(lowest Y) \hii regions. 

\subsection{The Systematic Uncertainty In $Y_{\rm P}$: A 
Semi-Empirical Approach}

Many observers have identified numerous sources of uncertainty affecting  
HII region helium abundance determinations (see, e.g., Davidson \& Kinman 
1985; Dinerstein \& Shields 1986; PSTE; S; ITL; Peimbert 1996).  Peimbert 
(1996) divides the errors associated with the determination of the 
primordial helium abundance into three groups: (I) errors in the 
determination of the line ratios; (II) errors in the interpretation of the 
line ratios; and (III) errors in the extrapolation to $Z$ $=$ 0.  Here we 
try to infer a reasonable estimate of the {\it systematic} uncertainty in 
Y$_{\rm P}$ by inspecting the various possible systematic effects in each 
group.

In Group I the errors in determining line ratios can be attributed
to measurements of the line ratios (including signal-to-noise in the 
line and sky subtraction), detector calibration, reddening corrections, 
and lack of corrections for possible underlying stellar absorption.  All 
of the effects in Group I have been discussed in detail in Skillman et 
al.\ (1994) and in previous studies.  To summarize, if detectors which 
have been tested for linearity (CCDs) are used, if several standard stars 
which have been previously observed with linear detectors (preferably the 
HST standards of Oke) are observed, if the targets are restricted to those 
objects of high excitation and high Balmer line equivalent width, and if 
one accumulates in excess of 10,000 photons in each of the He lines used, 
then it is possible to achieve an accuracy of 2\% in the relevant He/H 
line ratios.  Then, of all the effects described above, only unaccounted-for 
underlying stellar absorption would cause a systematic error, leading to 
an underestimate of the He abundance.  However, the presence or absence 
of this effect can be probed by measuring different He lines of different 
equivalent widths.  The general agreement between the different lines, in 
those cases with careful tests, indicates that the effect of underlying 
stellar absorption is of order 1\% or less.

In Group II the errors in the interpretation of line ratios can be 
attributed to correcting for the presence of neutral He, variations
in temperature structure (``temperature fluctuations"), the accuracy 
of the atomic data, the correction for the collisional excitation of 
HeI lines (primarily from the meta-stable triplet 2S level), correction for 
radiative transfer effects, and correction for collisional excitation 
of the HI lines (from the ground state). Taken in order:

(i) In principle, the presence of neutral He would systematically lower 
the observed He abundance. However, none of the tests performed so far 
have found any evidence of neutral He (see Skillman et al.\ 1994), and 
photoionization models indicate that it is not likely to be a problem 
for the objects included in these studies (see also the discussion in 
Vilchez \& Pagel (1988) and in PSTE). 

(ii) If there are variations in the electron temperature in the gas, 
the heavy element abundances derived from the collisionally excited 
lines would be underestimated and the He abundances would be slightly 
overestimated (due to the weak dependence of the He lines on the electron 
temperature, the effect is 1.5\% for $\lambda$5876 and $\lambda$6678 for 
an error of 1000K at 15,000K and about half of that for $\lambda$4471). 
Temperature variations appear to be much more likely at higher 
metallicities, but if supernovae are an important heating source in the 
\hii regions, then temperature fluctuations may be important
(Skillman 1995, Peimbert \etal 1991). 

(iii) The comparison of Smits (1996) with Brocklehurst (1972) would 
indicate that the atomic data and calculations of the recombination
emissivities for the He lines of interest are good to better than 1\%. 
Note, however, that there are much larger differences in the infrared 
transitions, and more work (both theoretical and observational) is 
desirable in this area.

(iv) The recent work by Kingdon \& Ferland (1995) gives us confidence 
that we are able to correct accurately for collisional excitation of HeI.
These corrections are usually of order 1 -- 3\%.  Not correcting would
systematically overestimate the He abundance.  The main problem here
is to determine the density sufficiently accurately.  ITL have argued 
that densities derived from [S~II] emission lines are not appropriate.
Since the different He~I emission lines have different 
density dependences, by using 
several lines it is possible to solve for the density, and thus the 
correction (this is essentially how ITL propose to 
solve for the electron density).
In general, the densities derived in this manner by ITL agree well with
those derived using the [S~II] lines.

(v) Since the work of Robbins (1968), it has been generally agreed that 
radiative transfer effects are unimportantly small for the bright lines 
which are used to derive the He abundances, particularly the singlet lines.  
Based on the results of photoionization modeling, in which the effects of 
collisional coupling of the singlets and triplets, radiative transfer 
effects, and collisional excitation were all treated simultaneously, 
Sasselov \& Goldwirth (1995) claimed that the He/H line ratios lead to 
systematic underestimates.  However, no He/H line ratios were presented 
in their paper.  Until such effects on the He/H line ratios are identified
and quantified, it seems reasonable to ignore this claim. 

(vi) Davidson \& Kinman (1985) showed that at the high temperatures found
in the lowest metallicity HII regions, collisional excitation of the HI
lines may be important.  Skillman \& Kennicutt (1993) showed that this 
effect, which is 
dependent on the neutral hydrogen fraction, is not significant for 
neutral H fractions less than 0.0001.  Straightforward calculation of the 
photoionization balance in an \hii region usually results in neutral H 
fractions less than this.  Photoionization codes often produce higher 
neutral H fractions, but this may be due to the approximations made
in the treatment of the ionizing radiation field.  This could be a very
difficult uncertainty to pin down, since the geometry of the gas 
distribution relative to the ionizing source has a strong influence on 
the neutral H fraction.  If this effect were important, it would result 
in an underestimate of the He abundance.  Perhaps 2\% represents a 
reasonable upper limit on the uncertainty of this effect.  

Group III errors are concerned with the extrapolation of the observed 
helium abundances to zero metallicity.  The ``classical" approach (Peimbert
\& Torres-Peimbert 1974) has been to fit the observations with a linear
Y versus Z relation (where the oxygen abundance usually serves as a
surrogate for Z) and to extrapolate to Z = 0.  For observations of low
metallicity \hii regions this linear fit may be thought of as the lowest
order contribution to a more general Y(Z) relation.  For the set B(C),
this extrapolation from the lowest metallicity data
is quite small: $\Delta$Y = 0.002$\pm$0.001(0.003$\pm$
0.001).  Since He/H is only expected to increase with Z, it is unlikely 
that this approach can systematically underestimate Y$_{\rm P}$.  While 
this linear fit may yield an upper bound on $Y_{\rm P}$, it does not 
necessarily provide a lower bound.  Indeed, as our second cut set (set C) 
shows, the Y versus Z slope appears to steepen at the very lowest 
metallicities.  If the assumption of linearity is relaxed, then, in 
principle, $Y_{\rm P}$ can be significantly lower.  

Given the different production sites of He and O, He/H may not be expected 
to track O/H well.  Indeed, Steigman, Gallagher \& Schramm (1989) suggested
that helium may correlate better with nitrogen and/or carbon than with
oxygen.  The observation by Pagel et al.\ (1992) that the dispersion in 
the He vs.\ N regression is less than that of the He vs.\ O regression 
lends some support to this expectation.  However, the observed linearity 
of Y with Z (where Z $\approx 20(O/H)$) over more than a decade in Z 
(e.g., OS) may reflect a balance between losses due to galactic winds 
(most important for low mass, low metallicity systems) and the metallicity 
dependence of the yields (O yield decreasing with increasing metallicity 
due to the increasing importance of stellar winds in the massive stars). 
Future work on accurate relative abundances to search for a ``second 
parameter" in the Y vs. Z relationship would be of great value. 

Since the data are entirely consistent with a linear Y vs. Z relation,
the uncertainty in the intercept should be of order the uncertainties in 
the best measured points at low metallicities (i.e., 2 -- 3\%).  Calculating 
the uncertainty in the intercept depends on knowledge of the source of the 
scatter in Y at a given Z, which presently is dominated by measurement 
uncertainties.

In the analysis presented here we have, to some extent, avoided the issue
of the extrapolation to zero metallicity by considering the helium abundance
determined from the best observed \hii regions (Y$_{\rm P} \leq$ Y$_{\rm 
OBS}$).  From five independent observations of IZw18 we found, Y$_{\rm P} 
\leq 0.230\pm$0.004; from the 13-14 \hii regions with the lowest helium 
abundances we derived, Y$_{\rm P} \leq 0.230\pm$0.003.

Finally, there remains the important question of how to combine different
sources of systematic uncertainties.  Since the possible errors are not
correlated, it makes no sense to add (linearly) all imaginable 
systematic errors to obtain an estimate of the total systematic error. 
Many potential error sources can be classified as unlikely, with bounds
constrained observationally.  Therefore it is even more unlikely that a 
single data point (let alone all of them) would suffer from more than one 
of the potential systematic errors at the amplitude of the observationally 
constrained limits.  The salient point is that all imaginable systematic 
errors appear to be limited to about 2 percent or less.  Thus, it seems 
reasonable to adopt as an estimate for the overall systematic error, 0.005, 
as proposed by PSTE.

We can also attempt to exploit the data itself to provide a bound on the 
size of possible systematic sources of error.  Although many of the 
sources of potential errors listed above might shift $Y$ and/or O/H in 
a systematic fashion, modifying the intercept and/or slope inferred 
from the Y versus O/H relation, their variation from source to source 
and observer to observer (telescope/detector to telescope/detector) 
would also contribute to the overall dispersion of the data around the 
``true" Y versus O/H relation.  We have therefore taken our best fit 
for set B (see Table 1) and examined the residuals, $Y - Y(B)$, as a 
function of O/H.  For the variance of the residuals we find 0.007; 
this should be compared to the error estimates for individual \hii 
region Y determinations which are, on average, 0.010.  It appears 
that the observers have been generous, perhaps overly generous, in 
their error estimates as first pointed out by PTSE.  
Indeed, this was already suggested by the 
small values of the reduced $\chi^2$ seen in Table 1.  We stress that 
since the variance of the residuals is small, there is no real statistical 
significance to the ITL claims that S96/KF improves the scatter.

As a further check on the stability of our $Y_{\rm P}$ estimates and 
to constrain many of the sources of possible systematic error, we 
have performed 40,000 runs of a statistical bootstrap (Olive \& Scully 
1995) using all the data (set A) with, and without the error estimates 
(we thank Tim Beers for suggesting this test to us and Sean Scully for 
doing the runs). In Figure 6 we show the resulting distributions for 
$Y_{\rm P}$ (when errors are included).  In both cases the distributions 
are closely Gaussian with $Y_{\rm P} = 0.234 \pm 0.002$ and 95\% CL upper 
bounds $\lsim$0.238.  This test suggests that unless {\bf ALL} values of 
Y should be systematically shifted (e.g., due to inaccurate atomic data), 
0.238 might provide a good upper bound to $Y_{\rm P}$, {\bf including} 
systematic errors.  However, to err on the side of conservatism, instead 
of an upper bound of 0.238 (including systematic uncertainties), we will 
adopt the set C(B) results in the subsequent discussion,
\beq
Y_{\rm P} = 0.230 \pm 0.003 \pm 0.005 (0.234 \pm 0.002 \pm 
0.005)
\eeq
where 0.003(0.002) represents the (Gaussian) statistical error 
($Y_{P}^{2\sigma} \leq 0.237$(0.239)) and 0.005 is a possible 
systematic 
offset in $Y_{\rm P}$ (leading to  ``firm" 2$\sigma$ upper bounds to 
$Y_{\rm P}$ of 0.242 or 0.244).  We will also explore the 
consequences 
of a larger value for $\Delta Y_{\rm sys}$.

\section{Discussion}

First let us ignore any possible systematic uncertainty in our adopted 
value of $Y_{\rm P}$ (eq. 5) to identify the range in the nucleon 
abundance ($\eta_{10}$) which follows from SBBN (including uncertainties 
in the neutron lifetime and any relevant nuclear reaction rates; Hata et 
al. 1995).  For $Y_{\rm P} = 0.234 \pm 0.002$,
\beq
\eta_{10} = 1.8 \pm 0.3
\label{eta}
\eeq
For an upper bound of $Y_{\rm P} \leq$ 0.239(0.244), the corresponding 
95\% CL upper bound on $\eta_{10}$ is 2.4(3.8).   Furthermore, the very 
low value of $\eta$ in (\ref{eta}), which is derived directly from \he4,
corresponds to a very low universal density of baryons,
\beq
\Omega_{\rm B} h^2 = 0.007 \pm 0.001
\eeq
Even for $Y_{\rm P} \leq 0.239(0.244)$, $\Omega_{\rm B} h^2 \leq$ 
0.009(0.014).
For the lower value of $Y_{\rm P} = 0.230 \pm 0.003$,
\beq
\eta_{10} = 1.4 \pm 0.3
\label{eta2}
\eeq
In this case for $Y_{\rm P} \leq$ 0.237(0.242), the upper bound on 
$\eta_{10}$ is 2.1(3.2).  Consequently, the density of baryons is even 
lower,
\beq
\Omega_{\rm B} h^2 = 0.005 \pm 0.001
\eeq
and for $Y_{\rm P} \leq 0.237(0.242)$, $\Omega_{\rm B} h^2 \leq$ 
0.008(0.012).

Although these estimates for $\eta$ are based solely on \he4,
concordance of SBBN with the observational data requires that the same 
value of $\eta$ yield acceptable abundances for the other light elements. 
For example, in a likelihood analysis based on \he4 and \li7 (the latter 
inferred from the metal-poor halo stars), Fields \& Olive (1996) and 
Fields \etal (1996) found consistency for a similarly low value of $\eta$.  
Using $Y_P = 0.234 \pm 0.003$, they found $\eta_{10} = 1.8$ as the most 
likely value and a 95\% CL range from 1.4 to 4.3.  Repeating this \he4/\li7 
analysis for the lower value of $Y_P = 0.230 \pm 0.003$ we find for the 
best fit value $\eta_{10} = 1.7$ and a 95\% CL range from 1.3 to 4.0.

 For such low nucleon abundances there is consistency (Dar 1995) between 
the SBBN predictions and the primordial abundances not only of \he4 and
$^7$Li, but also with the deuterium abundance as determined from observations 
of some QSO absorption line systems (Songaila et al. 1994; Carswell et al. 
1994; Rugers \& Hogan 1996).  The very high D abundance does pose a 
challenge to Galactic evolution models since it requires that the Galaxy 
has destroyed more than 90\% of its initial deuterium.  To do so while 
avoiding the overproduction of heavy elements may require the presence 
of supernovae driven Galactic winds (Scully \etal 1996).  When the 
likelihood analysis is extended to include the high primordial D/H, 
Fields \etal (1996) find that the peak value of $\eta_{10}$ is 1.7 
with a 95\% CL range between 1.5 and 2.4; we find that this range remains 
essentially the same for either $Y_P = 0.234$ or 0.230.  The corresponding 
value for the combination $\Omega_{\rm B} h^2$ is now 0.006 with a 95\% CL 
range between 0.005 and 0.009.

Alternatively we may use either of the two recent determinations of 
primordial deuterium in high redshift, low metallicity QSO
absorption systems to pin down $\eta$ and the corresponding range  
for $Y_{\rm P}$ predicted by SBBN which we may then compare to 
our adopted value (range) for $Y_{\rm P}$.

\subsection{High-D}

If the high abundance of deuterium derived from the observations of 
Songaila et al. (1994), Carswell et al. (1994) and Rugers and Hogan 
(1996) is representative of the primordial D abundance, then $1.3 
\leq \eta_{10} \leq 2.7$ and 0.231 $\leq Y_{\rm SBBN} \leq$ 0.239 
(95\% CL) (Hata et al. 1996).  This range, as already seen above, is in 
excellent agreement with our adopted range for Y$_{\rm P}$ (eq. 5) 
derived from the data and may be used to infer a restrictive upper 
bound to the effective number of equivalent light neutrinos 
($\Delta N_{\nu} \equiv  N_{\nu} - 3$).  For a systematic offset to 
$Y_{\rm P}$ of $\Delta Y_{\rm sys}$ = 0(0.005),
\beq
\Delta N_{\nu} \leq 0.5(0.8)
\eeq
Notice that if $Y_{\rm P}$ $>$ 0.239, $N_{\nu}$ $>$ 3.0 would be 
required.  The 95\% CL upper limit on the number of light degrees 
of freedom from the likelihood analysis  of Fields \etal (1996) 
based on \he4 and \li7 is (Olive \& Thomas 1996) $\Delta N_\nu < 1.0$ 
for $Y_P = 0.234 \pm 0.002$ and $\Delta N_\nu < 0.7$ for $Y_P = 
0.230 \pm 0.003$ (in both cases $\sigma_{\rm sys} = 0.005$ was 
assumed).

\subsection{Low-D}

In contrast, if the deuterium abundances derived for two different 
lines-of-sight from the data of Tytler, Fan \& Burles (1996) and of 
Burles \& Tytler (1996) provide good estimates of the true primordial 
value, then $5.1 \leq \eta_{10} \leq 8.2$ and 0.246 $\leq Y_{\rm SBBN} 
\leq$ 0.252 (95\% CL) (Hata et al. 1996).  Consistency with SBBN 
($N_{\nu} \geq$ 3.0) can only be recovered if systematic effects in 
deriving $Y_{\rm P}$ from the data have led to an underestimate by an 
amount $\Delta Y_{\rm sys} \geq$ 0.009 (i.e., consistency would require 
that $Y_{\rm P} \geq$ 0.246 compared to our upper bound of 0.237).  
Alternatively, if $\Delta Y_{\rm sys} \leq$ 0.005, $N_{\nu} \leq$ 2.7.

\section{Summary}

For nearly two decades low metallicity, extragalactic \hii regions 
have been studied as a probe of the primordial abundance of helium.  
{}From observations of ten such regions Lequeux et al. (1979) derived 
$Y_{\rm P} = 0.233 \pm 0.005$.  Using four carefully studied \hii 
regions, Torres-Peimbert et al. (1989) found $Y_{\rm P} = 0.230 \pm 0.006$, 
establishing the competitiveness of quality with quantity.  On the basis 
of nineteen extragalactic \hii regions PSTE inferred $Y_{\rm P} = 0.228 
\pm 0.005$, and building on this data set OS added the data of S to find 
for 41 (21) low metallicity \hii regions $Y_{\rm P} = 0.232 \pm $0.003 
(0.229 $\pm$ 0.005).  In this paper we have considered the new data from 
ITL which, at first glance, seems to yield a much larger value for 
$Y_{\rm P}$.  In contrast, we have found that the ITL data are fully 
consistent with those of PSTE and S and therefore we have combined these 
sets in an analysis of 62 (32) \hii regions.  From the Y versus O/H 
correlation we derive (see Table 1) $Y_{\rm P} = 0.234 \pm 0.002$ 
(0.230 $\pm$ 0.003).  We have also considered the five independent 
observations (PSTE; Skillman \& Kennicutt 1993; ITL) of the two 
knots in IZw18 to derive an upper bound to $Y_{\rm P}$ of 0.230 
$\pm$ 0.004, and in an extension of this approach to $Y_{\rm P}$ we 
have considered the weighted mean (and the weighted mean plus 2-sigma) 
for the \hii regions with the lowest values of Y to derive 
$Y_{\rm P} \leq 0.230$ $\pm$ 0.003.  These results have led us to 
adopt a ``95\% CL" upper bound to $Y_{\rm P}$ of 0.237 (for set B 
with 62 \hii regions, this upper bound is 0.239).

The availability of large numbers of carefully observed, low 
metallicity \hii regions has permitted estimates of $Y_{\rm P}$ 
whose statistical uncertainties are very small ($\approx$ 1\%).  
However, there remains the possibility that in the process of using 
the observational data to derive the abundances, contamination by 
unacknowledged systematic errors has biased the inferred value of 
$Y_{\rm P}$.  The observers have identified many potential sources 
of such systematic errors (Davidson \& Kinman 1985; PSTE; S; ITL; 
Peimbert 1996) and, where possible, have designed their observing 
programs to minimize such uncertainties and/or to account for them.  
Here we have noted that many of the identified sources of potential 
systematic errors would vary from \hii region to \hii region and 
from observer (telescope/detector) to observer, introducing not only 
a systematic offset in the derived value of $Y_{\rm P}$, but also an
accompanying dispersion in the Y versus O/H relation.  The very small 
values of the reduced $\chi^{2}$ for our fits suggest that the observers' 
error estimates may already account for some sources of systematic error.  
We have performed several tests confirming this and conclude that our 
determinations of $Y_{\rm P}$ are robust in the absence of some yet 
to be identified systematic offset which shifts all the data uniformly.  
Nonetheless, in discussing the consequences of our derived value of 
$Y_{\rm P}$ for cosmology and for particle physics we have allowed 
for a possible systematic offset $\Delta Y_{\rm sys}$ = 0.005.

For SBBN the low value we derive for $Y_{\rm P}$, consistent with 
previous results (Lequeux et al. 1979; Torres-Peimbert et al. 1989; 
PSTE; OS), implies a low nucleon abundance but is entirely consistent 
with the inferred primordial abundances of $^7$Li and D (from the QSO 
absorbers studied by Rugers \& Hogan 1996).  Provided that the systematic 
error in $Y_{\rm P}$ is not large, there is a meaningful constraint on 
the effective number of equivalent light neutrinos (Steigman, Schramm 
\& Gunn 1977). If, instead, the low deuterium abundance inferred from 
the data of Tytler, Fan \& Burles (1996) and of Burles \& Tytler (1996) 
is the ``true" primordial value, there is a challenge to SBBN unless 
$\Delta Y_{\rm sys}$ is large (so that $Y_{\rm P} \geq$ 0.246).

\vskip 0.5truecm
\noindent {\bf Acknowledgments}
\vskip 0.5truecm
We owe a debt of gratitude to R.C. Kennicutt, E. Terlevich and R.J. 
Terlevich for sharing their unpublished data with us. We thank S. Scully 
for running the statistical bootstrap and for preparing Figures 3 and 6.
We also are pleased to thank T. Beers, D. Garnett, N. Hata, Y. Izotov, 
D. Kunth, B. Pagel, D. Schaerer, S. Viegas and T. Walker for very 
informative and valuable discussions.  The work of KAO is supported in 
part by DOE grant DE-FG02-94ER-40823 and that of GS by DOE grant 
DE-AC02-76ER-01545. EDS is grateful for partial support from a NASA 
LTSARP grant No. NAGW-3189 and a Bush Sabbatical Fellowship from the 
Graduate School of the University of Minnesota and would especially like to
thank the Max-Planck-Institute for Astrophysics for their hospitality
during his sabbatical year.  Parts of this work were carried out while KAO and 
GS were participants in the INT Workshop on Nucleosynthesis in Stars, 
Supernovae and the Universe and also while GS was a visitor at the 
Instituto Astronomico e Geofisico of the University of Sao Paulo (Brasil) 
and they are grateful for the hospitality provided.
\newpage
\beginapjbib

\bibitem Berrington, J.A. \& Kingston, A.E. 1987, J. Phys. B, 20, 6631

\bibitem Bevington, P.R. \& Robinson, D.K. 1992 Data Reduction and 
Error
 Analysis for the Physical Sciences (New York: McGraw Hill)

\bibitem Brocklehurst, M. 1972, MNRAS, 157, 211 (B72)

\bibitem Burles, S. \& Tytler, D. 1996, ApJ, 460, 584

\bibitem Carswell, R.F., Rauch, M., Weymann, R.J., Cooke, A.J. \&
Webb, J.K. 1994, MNRAS, 268, L1

\bibitem Clegg, R.E.S. 1987, MNRAS, 229, 31P

\bibitem Dar, A. 1995, ApJ, 449, 550

\bibitem Davidson, K. \& Kinman, T.D. 1985, ApJS, 58, 321
\bibitem Dinerstein, H.L. \& Shields, G.A. 1986, ApJ, 311, 45

\bibitem Ferland, G.J. 1986, ApJ, 310, L67

\bibitem Fields, B.D. 1996, ApJ, 456, 478 

\bibitem Fields, B.D. \& Olive, K.A. 1996, Phys Lett B368, 103

\bibitem Fields, B.D., Kainulainen, K., Olive, K.A., \& Thomas, D. 1996
New Astronomy, 1, 77

\bibitem Hata, N., Scherrer, R.J., Steigman, G., Thomas, D., \&
Walker, T.P. 1995,  Phys. Rev. Lett., 75, 3977

\bibitem Hata, N., Steigman, G., Bludman, S. \& Langacker, P. 1996, 
Phys. Rev.
D, in press, astro-ph/9603087

\bibitem Izotov, Y.I., Thuan, T.X., \& Lipovetsky, V.A. 1994
ApJ 435, 647. (ITL)

\bibitem Izotov, Y.I., Thuan, T.X., \& Lipovetsky, V.A. 1996, (ITL)

\bibitem Kingdon, J. \& Ferland, G.J. 1995, ApJ, 442, 714 (KF)


\bibitem Kobulnicky, H.A. \& Skillman, E.D. 1996, ApJ, 471, 211

\bibitem Kunth, D., Lequeux, J., Sargent, W.L.W., \& Viallefond, F. 
1994, A\&A, 282, 709

\bibitem Kunth, D. \& Sargent, W. 1983, ApJ, 273, 81

\bibitem Lequeux, J., et. al. 1979, A \& A, 80, 155

\bibitem Oke, J. B. 1990, AJ, 99, 1621

\bibitem Olive, K.A. \& Scully, S.T. 1996, IJMPA, 11, 409

\bibitem Olive, K.A., Steigman, G. \& Walker, T.P. 1991, ApJ, 380, L1

\bibitem Olive, K.A., \& Steigman, G. 1995, ApJS, 97, 49

\bibitem Olive, K.A. \& Thomas, D. 1996, preprint in preparation

\bibitem Pagel, B.E.J. \& Kazlauskis, A. 1992, MNRAS, 256, 49P

\bibitem Pagel, B E.J., Simonson, E.A., Terlevich, R.J.
\& Edmunds, M. 1992, MNRAS, 255, 325 (PSTE)

\bibitem  Pagel, B.E.J., Terlevich, R.J. \& Melnick, J. 1986,
 PASP, 98, 1005 (PTM)

\bibitem Peimbert, M. 1996, Rev. Mex. Astr. Astrofis., in press

\bibitem Peimbert, M. Sarmiento, A. \& Fierro, J. 1991, PASP, 103, 815

\bibitem Peimbert, M. \& Torres-Peimbert, S. 1974 Ap J, 193, 327

\bibitem Robbins, R.R. 1968, ApJ, 151, 511

\bibitem Rugers, M. \& Hogan, C. 1996, ApJ,  259, L1

\bibitem Sasselov, D. \& Goldwirth, D.S. 1995, ApJ, 444, L5

\bibitem Sawey, P.M.J. \& Berrington, K.A. 1993, Atomic Data Nucl. Data 
Tables, 55, 81

\bibitem Scully, S.T., Cass\'{e}, M., Olive, K.A.,  
 \& Vangioni-Flam, E. 1996, ApJ, in press, astro-ph/9607106

\bibitem Searle, L. \& Sargent, W.L.W. 1971, ApJ, 173, 25

\bibitem Simonson, E. A. 1990, Ph. D. thesis, University of Sussex

\bibitem Skillman, E. 1985, ApJ, 290, 449

\bibitem Skillman, E., \& Kennicutt 1993, ApJ, 411, 655 (S)

\bibitem Skillman, E., Terlevich, R.J., Kennicutt, R.C., Garnett, D.R.,
\& Terlevich, E. 1994, ApJ, 431, 172 (S)

\bibitem Skillman, E., et al. 1996, ApJ Lett (in preparation) (S)

\bibitem Smits, D.P. 1991a, MNRAS, 248, 193

\bibitem Smits, D.P. 1991b, MNRAS, 251, 316

\bibitem Smits, D.P. 1996, MNRAS, 278, 683 (S96)

\bibitem Songaila, A., Cowie, L.L., Hogan, C. \& Rugers, M.,
1994, Nature, 368, 599

\bibitem Steigman, G., Gallagher, J.S. \& Schramm, D.N. 1989, Comments on 
Astrophysics, Vol.~XIV, p.~97

\bibitem Steigman, G., Schramm, D.N. \& Gunn, J. 1977, Phys. Lett., 
B66, 202

\bibitem Torres-Peimbert, S., Peimbert, M., \& Fierro,
   J. 1989, ApJ, 345, 186

\bibitem Tytler, D., Fan, X.-M., and Burles, S. 1996, Nature, 381, 207

\bibitem Vilchez, J.M. \& Pagel, B.E.J. 1988, MNRAS, 231, 257

\bibitem Walker, T. P., Steigman, G., Schramm, D. N., Olive, K. A.,
\& Kang, H. 1991, ApJ, 376, 51 (WSSOK)

\endapjbib

\newpage
\noindent{\bf{Figure Captions}}

\vskip.3truein

\begin{itemize}

\item[]
\begin{enumerate}
\item[]
\begin{enumerate}

\item[{\bf  Figure 1:}] A comparison of the calculations of the collisionally
excited emission to recombination emission rates (C/R) for four
HeI recombination lines using the formulae of Clegg (1987) and
Kingdon \& Ferland (1995).  The calculations were carried out
for an electron density of 100, which is appropriate for most
low metallicity HII regions.  The large difference seen for the
$\lambda$ 7065 line arises because Clegg (1987) used the recombination
coefficients from Brocklehurst (1972) which, for $\lambda$ 7065,
have been shown to be in error by Smits (1991a,b).

\item[{\bf Figure 2:}]  The helium (Y) and oxygen (O/H) abundances 
for the extragalactic \hii regions of the 1st cut data sets (OS) from 
PTSE and S (open circles), and from ITL (filled circles).  Regions 
excluded by ITL are shown as crossed circles.  Lines connect the 
same regions observed by different groups. 

\item[{\bf  Figure 3:}]   The distribution of the error weighted 
determinations of $Y_{\rm P}$ (open circles) from a 40,000 run 
statistical bootstrap using the modified OS data set (excluding the 
four points with lowest O/H).  The solid curve is the best fit gaussian. 

\item[{\bf  Figure 4:}]  The residuals, $Y - \langle Y \rangle$, of the 
data (Y) compared to the weighted mean ($\langle Y \rangle$) versus 
the oxygen abundance for the first cut set B.
 
\item[{\bf  Figure 5:}]   The running average (weighted means) of the 
\he4 abundance, $Y$, 
for the first $N$ (lowest $Y$) \hii regions.  Also shown are the 
2$\sigma$ bounds to the weighted means. 

\item[{\bf  Figure 6:}]   The distribution of the error weighted 
determinations of $Y_{\rm P}$ (open circles) from a 40,000 run 
statistical bootstrap using the full data set (A).  The solid curve is the 
best fit gaussian. 

\end{enumerate}
\end{enumerate}
\end{itemize}

\newpage

\begin{figure}[htb]
\hspace{+1truecm}
\epsfysize=7.5truein
\epsfbox{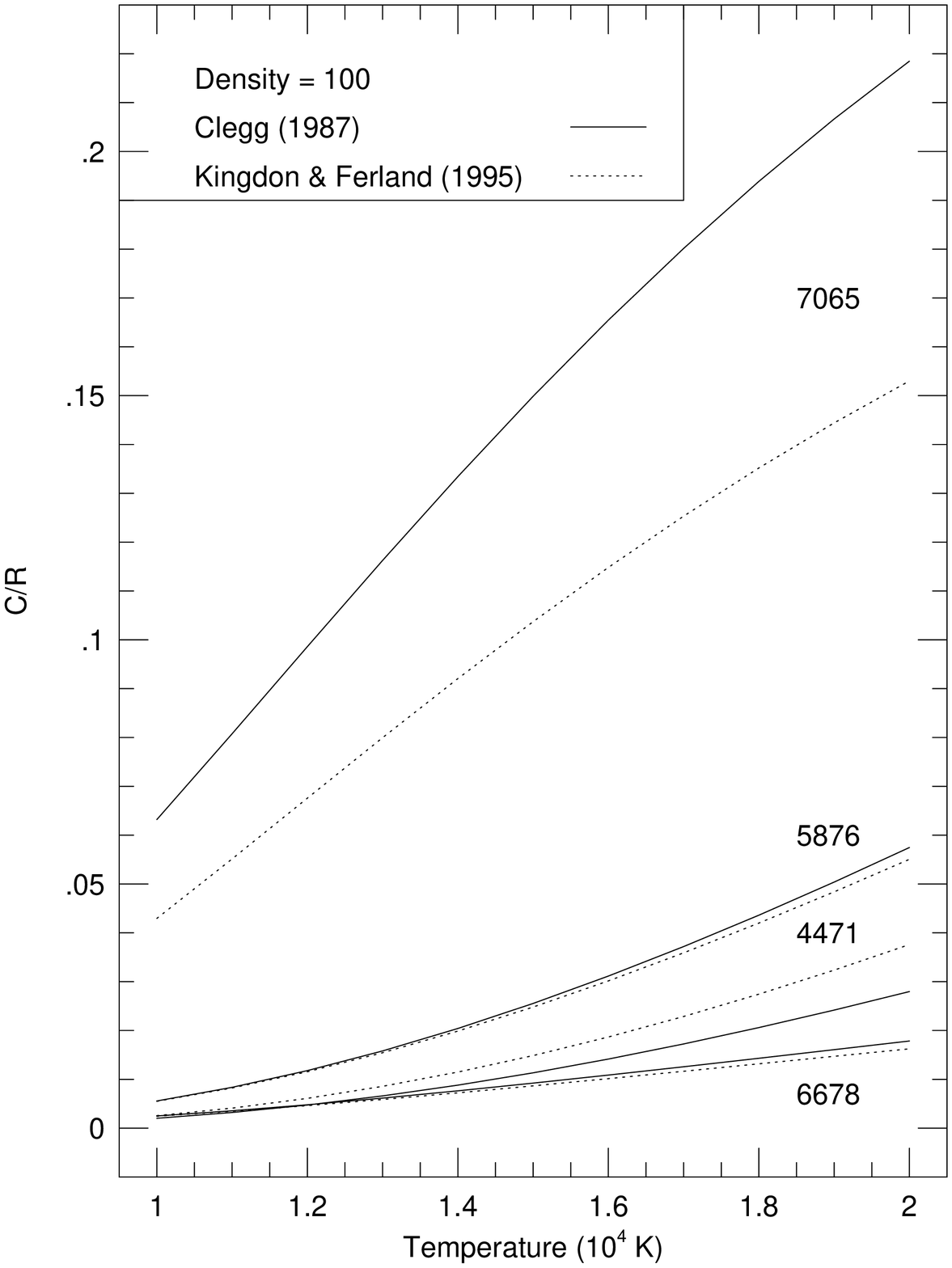}
\end{figure}  

\newpage

\begin{figure}[htb]
\hspace{+1truecm}
\epsfysize=8.0truein
\epsfbox{fig2.epsf}
\end{figure}  

\newpage

\begin{figure}[htb]
\epsfysize=8.0truein
\epsfbox{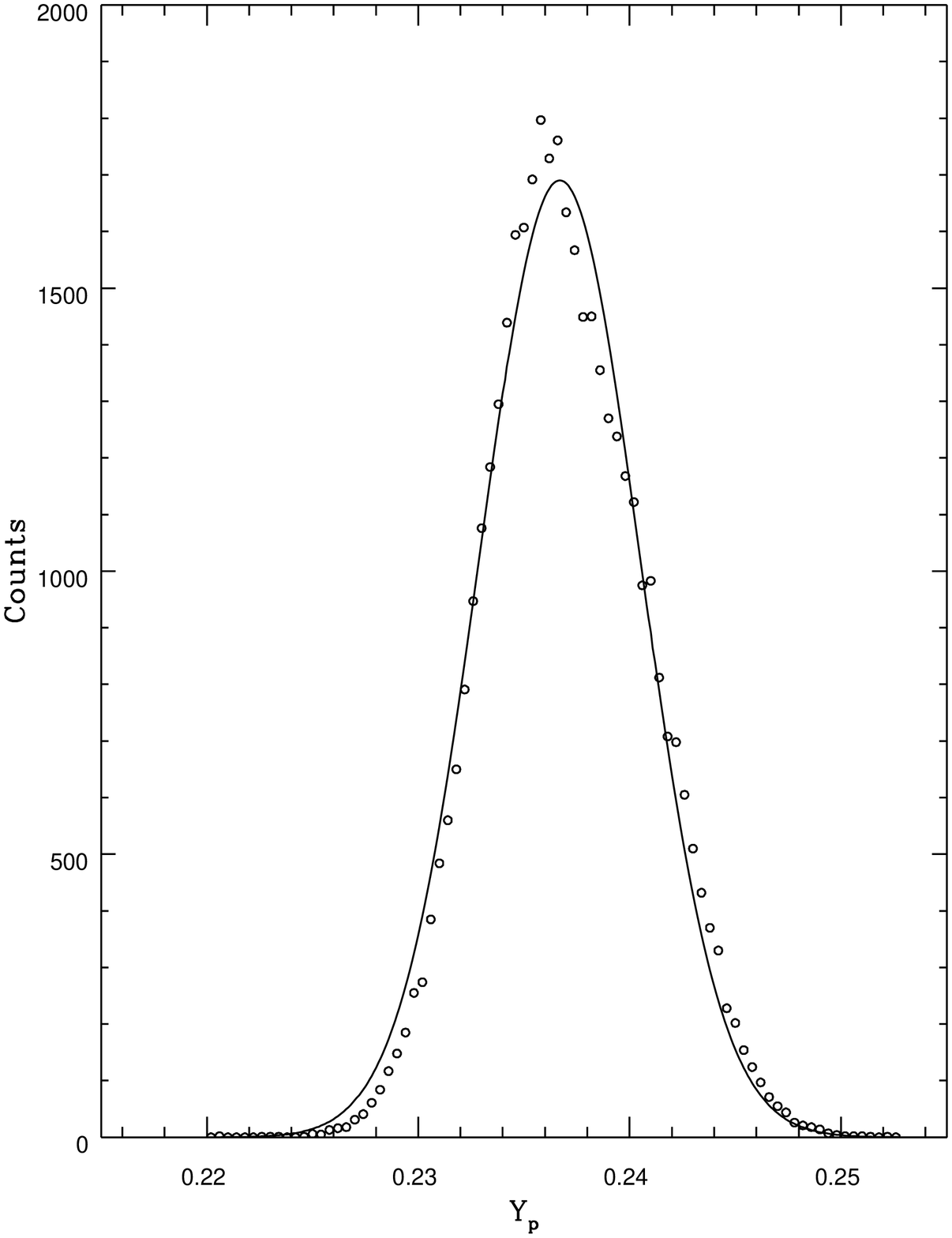}
\end{figure}  

\newpage

\begin{figure}[htb]
\hspace{+1truecm}
\epsfysize=7.5truein
\epsfbox{fig4.epsf}
\end{figure}  

\newpage

\begin{figure}[htb]
\hspace{+1truecm}
\epsfysize=7.0truein
\epsfbox{fig5.epsf}
\end{figure}  

\newpage

\begin{figure}[htb]
\hspace{-1truecm}
\epsfysize=8.0truein
\epsfbox{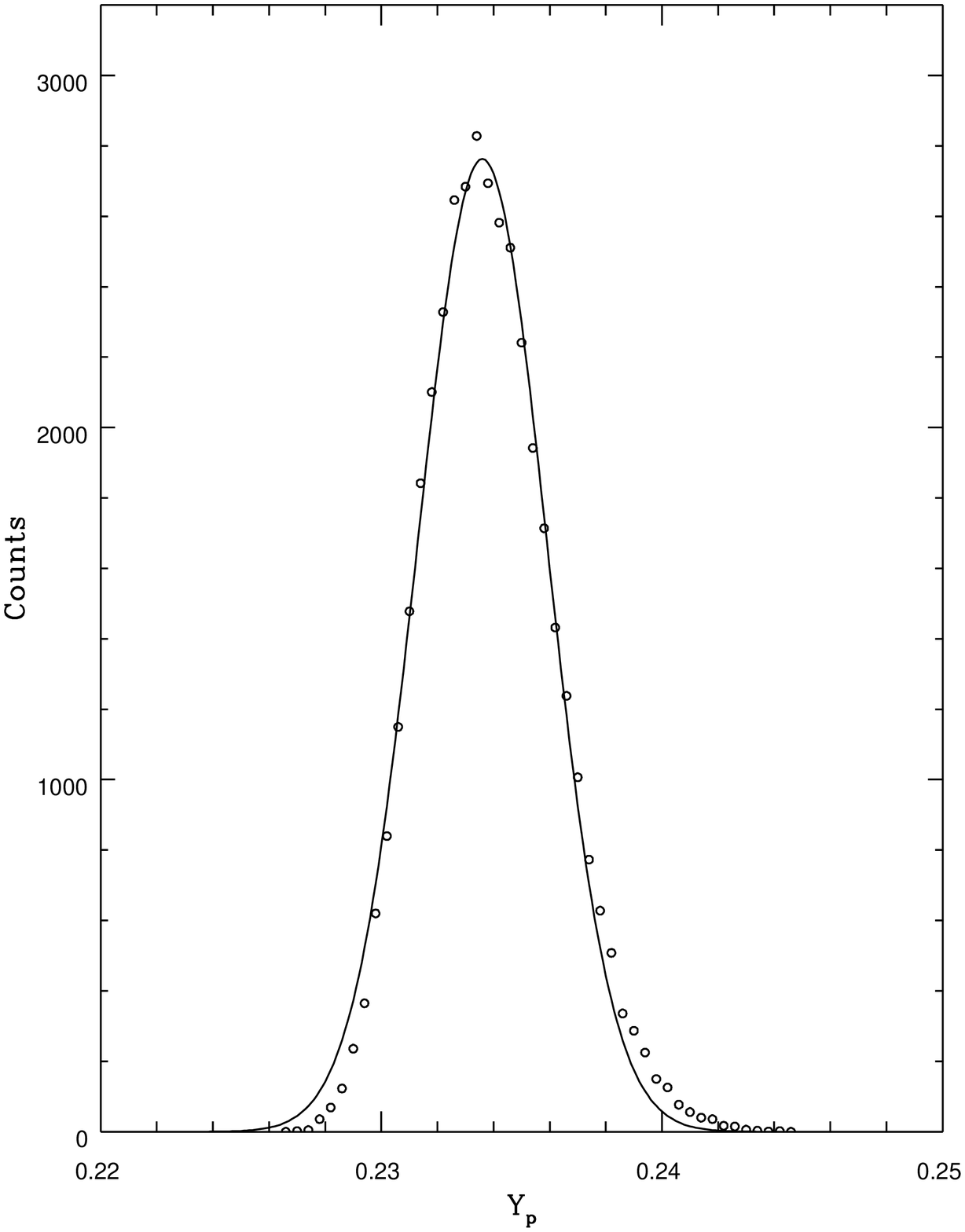}
\end{figure}

\end{document}